\documentclass{article}%
\usepackage{amsmath}
\usepackage{amsfonts}
\usepackage{amssymb}
\usepackage{graphicx}%
\newcommand{\bpartial}{\mathop{\partial\kern -4pt\raisebox{.8pt}{$|$}}}
\newcommand{\bra}{\mathopen{[\kern-1.6pt[}}
\newcommand{\ket}{\mathclose{]\kern-1.5pt]}}
\newcommand{\bbra}{\mathopen{[\kern-2.2pt[\kern-2.3pt[}}
\newcommand{\bket}{\mathclose{]\kern-2.1pt]\kern-2.3pt]}}

\makeindex
\begin{document}
        \title {\large{ \bf String gravitational equations with Hermitian structure }}

        \vspace{3mm}

        \author {  \small{ \bf  F. Naderi}\hspace{-1mm}{ \footnote{ e-mail: f.naderi@azaruniv.edu}} ,{ \small
                } \small{ \bf  A. Rezaei-Aghdam}\hspace{-1mm}{
                \footnote{Corresponding author. e-mail:
                        rezaei-a@azaruniv.edu}} \,and \small{ \bf  F. Darabi}\hspace{-1mm}{
                \footnote{e-mail:
                        f.darabi@azaruniv.edu}}\\
        {\small{\em Department of Physics, Faculty of Basic Sciences, Azarbaijan Shahid Madani University}}\\
        {\small{\em   53714-161, Tabriz, Iran  }}}

\maketitle

\begin{abstract}
        We consider a string model at one-loop related to a $\sigma$-model whose antisymmetric tensor field is constructed as complex structure on the background manifold, specially on a manifold $R\times N$ where $N$ is a complex manifold. As an example, we consider a homogeneous anisotropic $(1+4)$-dimensional $\sigma$-model where space part of the background is a $4$-dimensional complex manifold. By solving the related one-loop  $\beta$-functions we obtain a static  solution so that by reduction of this solution to $(1+3)$-dimension we obtain a static solution of Einstein equation where the matter sector is effectively interpreted as an inhomogeneous,  anisotropic and barotropic matter  satisfying all the  energy conditions. Finally, the $T$-dual background of the solution is investigated and it is shown that the duality transformation and reduction processes commute with each other.
        $$
        $$
        \textit{Keywords}: String gravitational model; $\sigma$-model; Hermitian structure.

        PACS numbers: $04.60.Cf$.
\end{abstract}
\section{Introduction}

String theory is the most consistent  theory for unifying  the
fundamental interactions, incorporating  gravity.
The cosmological implications of string theory (string cosmology) have recently received considerable attention \cite{1a,1b}. In this context, the string theory has been used in the study of the physical situation at very early universe \cite{string cosmology}.

The string theory in curved background, generally known as a conformal $\sigma$-model, is concerned with the massless modes of the string including metric, a rank two antisymmetric tensor field $B_{\mu\nu}$ and a dilaton field $\phi$. In the $\sigma$-model approach, the conformal invariance condition is the vanishing of the $\beta$-functions in all loops of the fields which may be interpreted as field equations of effective space-time action \cite{2,fradkin,beta}.

An interesting class of manifolds in the study of (super)string theory includes complex manifolds in which the set of solutions of string equations is based on complex non-K\"{a}hler manifolds \cite{cm1a,cm1b,cm2,cm3,cm4}. In \cite{witten}, Witten has constructed a topological $\sigma$-model where the complex structure $J$ of target space appeared in the Lagrangian in an interaction term like Kalb-Ramond field, through a supersymmetry. From another but related point of view, the flux $H$ could be related to the almost complex structure by imposing supersymmetry  as\footnote{Here $*$ is \textit{Hodge} star.}  $dJ\sim *H$\cite{ads,dj1,dj2}.

In this paper, we are interested in the idea of our recent paper to establish a physical interpretation for the complex structures \cite{rn}.
We know that in general relativity the gravitational effects are characterized by a geometrical structure, so called  metric, on a Riemannian manifold. But, on a complex
manifold there is an additional geometrical structure, namely the complex structure $J$, which may be capable of carrying physical significance. In this regard, we have
already presented a matter interpretation for the almost complex structure \cite{rn}. Here, following the same idea we will investigate a  $\sigma$-model where the role of antisymmetric  $B$-field tensor is played by the complex structure (or more clearly by the fundamental
antisymmetric K\"{a}hler form $\Omega$ which is related to the complex structure with a hermitian metric).  In this way, similar to the metric, the antisymmetric tensor
field in the
context of $\sigma$-model will be a geometrical structure of the manifold. We will assume that the background space-time is a $(1+4)$-dimensional manifold such that the
$4$-dimensional space part is a complex manifold carrying a complex structure. Then, the $B$-field defined on the space-time is considered to be related with the complex
structure of $4$-dimensional part. In this way, the kinetic term of complex structure will appear in the effective action. At this level, we can obtain
the solution of the lowest order of $\beta$-functions in the non-critical dimension with a non-zero central charge deficit.

The plane of the paper is as follows. In section $2$, we start by collecting some preliminaries about string effective gravitational equations and Hermitian structure. In section $3$, we introduce the special backgrounds of $(1+4)$-dimensional $\sigma$-model consisting of a hermitian metric and a complex structure as $B$-field. Then, the solution of $\beta$-functions are investigated. In section $4$, dimension reduction of the static $(1+4)$-dimensional solutions to $(1+3)$-dimensional space-time is performed. Then, we discuss the nature of derived  solutions, with the particular anisotropic and inhomogeneous form of energy momentum tensor given by dilaton and $B$-field. In section $5$, $T$-dual solutions are investigated. The paper ends with a conclusion.


\section{ Review on $\sigma$-model equations and Hermitian structure}
Consider the two dimensional $\sigma$-model in a curved background $M$, where its action is expressed by
\begin{eqnarray}\label{sigma}
I=\frac{1}{4\pi\alpha '}\int_{\Sigma}{\sqrt{h}d^2z(h^{\alpha\beta}\hat{g}_{\hat{\mu}\hat{\nu}}\partial_{\alpha}{X}^{\hat{\mu}}\partial_{\beta}{X}^{\hat{\nu}}+\epsilon^{\alpha\beta}\hat{B}_{\hat{\mu}\hat{\nu}}\partial_{\alpha}{X}^{\hat{\mu}}\partial_{\beta}{X}^{\hat{\nu}}+\alpha ' R^{(2)}\hat{\phi})},
\end{eqnarray}
such that the $h$ denotes the metric of worldsheet $\Sigma$, $\hat{g}_{\hat{\mu}\hat{\nu}}$ is the metric of $D$-dimensional space-time $M$, $\hat{\phi}$ and $\hat{B}_{\hat{\mu}\hat{\nu}}$ are dilaton and antisymmetric tensor known as Kalb-Ramond field on $M$, respectively. The $\hat{\mu},\hat{\nu}$  are $D$-dimensional target space-time indices and $\alpha,\beta$ are worldsheet indices and $R^{(2)}$ is the scalar curvature of the world sheet.

In the leading-order of  coupling $\alpha'$, the conditions of conformal (Weyl) invariance of $\sigma$-model \eqref{sigma}, i.e. the one-loop $\beta$-functions, are given by \cite{2,fradkin, beta}
\begin{eqnarray}\label{beta1}
\beta_{\hat{\mu}\hat{\nu}}^{~(\hat{g})}=\hat{R}_{\hat{\mu}\hat{\nu}}-\frac{1}{4}\hat{H}^{2}_{\hat{\mu}\hat{\nu}}-\nabla_{\hat{\mu}}\nabla_{\hat{\nu}}\hat{\phi},
\end{eqnarray}
\begin{eqnarray}\label{beta2}
\beta_{\hat{\mu}\hat{\nu}}^{~(\hat{B})}=\nabla^{\hat{\mu}}(e^{\hat{\phi}}\hat{H}_{\hat{\mu}\hat{\nu}\hat{\rho}}),
\end{eqnarray}
\begin{eqnarray}\label{beta3}
\beta^{~(\hat{\phi})}=-\hat{R}+\frac{1}{12}\hat{H}^{2}+2\nabla_{\hat{\mu}}\nabla^{\hat{\mu}}\hat{\phi}+(\partial_{\hat{\mu}}\hat{\phi})^{2}+\Lambda.
\end{eqnarray}
These equations are identical with the equations obtained by variation of the following  effective action with respect to the fields $\hat{g}_{\hat{\mu}\hat{\nu}}$, $\hat{B}_{\hat{\mu}\hat{\nu}}$ and $\hat{\phi}$, respectively \cite{2,fradkin,beta}
\begin{eqnarray}\label{effective}
S(\hat{g}_{\hat{\mu}\hat{\nu}},\hat{B}_{\hat{\mu}\hat{\nu}},\hat{\phi})=\frac{1}{2}\int d^{D}x  \sqrt{-\hat{g}}~e^{\hat{\phi}}(\hat{R}-\frac{1}{12}\hat{H}_{\hat{\mu}\hat{\nu}\hat{\rho}}\hat{H}^{\hat{\mu}\hat{\nu}\hat{\rho}}+\partial_{\hat{\mu}}\hat{\phi}\partial^{\hat{\mu}}\hat{\phi}-\Lambda).
\end{eqnarray}
Here, $\hat{H}^{2}_{\hat{\mu}\hat{\nu}}=\hat{H}_{\hat{\mu}\hat{\sigma}\hat{\rho}}\hat{H}_{\hat{\nu}}^{~\hat{\sigma}\hat{\rho}}$,  $\hat{H}^{2}=\hat{H}_{\hat{\mu}\hat{\nu}\hat{\rho}}\hat{H}^{\hat{\mu}\hat{\nu}\hat{\rho}}$, where the field strength $\hat{H}$ of $B$-field is defined as follows
\begin{eqnarray}\label{12}
\hat{H}_{\hat{\mu}\hat{\nu}\hat{\rho}}=\partial_{\hat{\mu}}\hat{B}_{\hat{\nu}\hat{\rho}}+\partial_{\hat{\nu}}\hat{B}_{\hat{\rho}\hat{\mu}}+\partial_{\hat{\rho}}\hat{B}_{\hat{\mu}\hat{\nu}}.
\end{eqnarray}
This action is so called  string effective action at one-loop.

In the non-critical theory, the constant $\Lambda$ is related to the central charge deficit of the original theory, i.e. $\Lambda=\frac{2(D-26)}{3\alpha'}$ \cite{landa, batakis1,batakis2}. It is analogous to the non-vanishing
cosmological constant in   cosmology in standard theory of gravity \cite{CC,cc2}. In this case, the one-loop $\beta$-functions requires some string scale curvature and this requires to consider higher-order corrections to the $\beta$-functions \eqref{beta1}-\eqref{beta3}.

Vanishing of the one-loop $\beta$-functions will provide us with a conformal $\sigma$-model at one-loop and  it is at the ground of the so-called \textit{string cosmology} at one-loop \cite{string cosmology}. It is worth mentioning that this expresses the lowest order conformality and dose not indicate that the $\sigma$-model is exactly-conformal.

The above formalism is in the \textit{string frame}. Another useful frame, in which the string effective action \eqref{effective} appears as the $D$-dimensional Einstein-Hilbert action of general relativity, is the \textit{Einstein frame}. The new metric is called Einstein metric and is obtained by a rescaling of the string metric as follows \cite{batakis1,batakis2}
\begin{eqnarray}
\tilde{\hat{g}}_{\hat{\mu}\hat{\nu}}=e^{\frac{2\hat{\phi}}{D-2}}\hat{g}_{\hat{\mu}\hat{\nu}}.
\end{eqnarray}
where, the $\tilde{\hat{g}}_{\hat{\mu}\hat{\nu}}$ denotes the $D$-dimensional Einstein metric whose Ricci scalar is $\tilde{\hat{R}}$. So, in the the Einstein frame the form of the action will be
\begin{eqnarray}
S=\frac{1}{2}\int d^{D}x  \sqrt{-\tilde{g}}~(\tilde{\hat{R}}-\frac{1}{12}e^{\frac{4\hat{\phi}}{D-2}}\hat{H}_{\hat{\mu}\hat{\nu}\hat{\rho}}\hat{H}^{\hat{\mu}\hat{\nu}\hat{\rho}}-\frac{1}{D-2}\partial_{\hat{\mu}}\hat{\phi}\partial^{\hat{\mu}}\hat{\phi}-e^{\frac{-2\hat{\phi}}{D-2}}\Lambda).
\end{eqnarray}
In this frame, the equations $\beta^{\{\hat{g},\hat{B},\hat{\phi}\}}=0$ in \eqref{beta1} - \eqref{beta3} are transformed to the Einstein equations as follows
\begin{eqnarray}\label{Ee}
\tilde{R}_{\hat{\mu}\hat{\nu}}-\frac{1}{2}\tilde{R}\tilde{g}_{\hat{\mu}\hat{\nu}}=\kappa^2(T^{(\hat{\phi})}_{\hat{\mu}\hat{\nu}}+T^{(\hat{H})}_{\hat{\mu}\hat{\nu}}),
\end{eqnarray}
where the energy-momentum tensors are given by
\begin{eqnarray}\label{t1}
\kappa^2T^{(\hat{\phi})}_{\hat{\mu}\hat{\nu}}=\frac{2}{D-2}\left(\partial_{\hat{\mu}}\hat{\phi}\partial_{\hat{\nu}}\hat{\phi}-\frac{1}{2}(\partial\hat{\phi})^2\tilde{g}_{\hat{\mu}\hat{\nu}}-\Lambda
e^{-\hat{\phi}}\tilde{g}_{\hat{\mu}\hat{\nu}}\right),
\end{eqnarray}
\begin{eqnarray}\label{t2}
\kappa^2T^{(\hat{H})}_{\hat{\mu}\hat{\nu}}=\frac{1}{4}\left(\hat{H}_{\hat{\mu}\hat{\kappa}\hat{\lambda}}\hat{H}_{\hat{\nu}}^{\hat{\kappa}\hat{\lambda}}-\frac{1}{6}\hat{H}^2\tilde{g}_{\hat{\mu}\hat{\nu}}\right).
\end{eqnarray}


\subsection{$\sigma$-model equations at one-loop with Hermitian structure}

We consider a  $\sigma$-model on a manifold $M$, at least one part of which is a complex manifold, so that the manifold carries another geometrical structure beside the metric, namely the complex structure $J$. Following the idea of giving a physical role to the complex structure, it is intriguing to investigate coupling of the complex structure to the string, included in the $B$-field interaction in the second term of the $\sigma$-model \eqref{sigma}.

Before attempting to investigate such a model, let us collect some preliminaries about the complex structure. Let $N$ be an \textit{even} dimensional manifold, an almost complex Hermitian structure $(g,J)$ on $N$ with coordinates $\{x^{i}\}$ consisting of a Reinmanian metric $g=g_{ij}dx^{i}\otimes dx^{j}$ and a structure $J=J_{i}^{~j}dx^{i}\otimes\partial_{j}$ satisfying \cite{nakahara}
\begin{eqnarray}\label{cs}
J^{~j}_{k}J^{~i}_{j}=-\delta^{i}_{k},
\end{eqnarray}
\begin{eqnarray}\label{hm}
g_{ij}J^{~i}_{m}J^{~j}_{n}=g_{mn},
\end{eqnarray}
which are definitions of the almost complex structure and Hermitian metric, respectively. These two conditions introduce a Hermitian structure. Fundamental two form or K\"{a}hler two form $\Omega$ is defined by $\Omega=g(.,J.)=\frac{1}{2}\Omega_{ij}dx^{i}\wedge dx^{j}$, whose components are given by $\Omega_{ij}=g_{ik}J^{~k}_{j}$. Then from {\eqref{cs}} and \eqref{hm}, we have
\begin{eqnarray}\label{3}
\Omega_{ij}=-\Omega_{ji}.
\end{eqnarray}
Nijenhuis tensor of an almost Hermitian manifold is defined by \cite{Nijenhuis1,Nijenhuis2}
\begin{eqnarray}\label{N}
N_{jk} ^{~i}=-J_{m}^{~i}\partial_{j}J_{k}^{~m}+J_{m}^{~i}\partial_{k}J_{j}^{~m}-J_{k}^{~m}\partial_{m}J_{j}^{~i}+J_{j}^{~m}\partial_{m}J_{k}^{~i}.
\end{eqnarray}
An almost complex structure $J$ on a manifold $M$ is \textit{integrable} if and only if $N_{jk} ^{~i}=0$; in this case the almost complex structure is called complex structure \cite{nakahara}.

As mentioned above, we want to describe a  $\sigma$-model on a special manifold where the $B$-field  is related to the complex structure in a special case. For this propose, an appropriate part of the $D$-dimensional manifold must be a complex manifold equipped with a complex structure $J$. On the other hand, noting the fact that the components of $J$ should be real themselves, through the equations \eqref{cs} and \eqref{hm}, a diagonal metric with Minkowski signature is not capable of being a Hermitian metric associated with a complex structure. This persuades us to consider a model on a manifold of the form $M=R\times N$, where $N$ is an \textit{even} dimensional manifold which is demanded to be a complex manifold. The manifold $N$ will describes the spatial part  and therefore the signature problem will be removed.
Summarizing, if the manifold $M$ is denoted by hatted indices  and $N$ by the $i$, $j$,... indices, we are going to consider an antisymmetric $B$-field of the form
\begin{eqnarray}\label{b}
\hat{B}= \frac{1}{2}\hat{B}_{\hat{\mu}\hat{\nu}}dx^{\hat{\mu}}\wedge dx^{\hat{\nu}}= \frac{\gamma}{2}~ \Omega_{ij}dx^{i}\wedge dx^{j},~~~~\hat{B}_{0i}=0,
\end{eqnarray}
where $\gamma$ is a coupling constant and $\Omega_{ij}=g_{jk}J_{i}^{~k}$ is the Hermitian K\"{a}hler two form associated to $g_{ij}$.

The basic ingredient of this model is therefore the dilaton and a complex structure in  the form of $B$-field. This ansatz for $B$-field has been already used in supersymmetry as a result of supersymmetry condition which imposes an ansatz on the dilaton field, too \cite{witten}. But, here we have used a different
idea in choosing this ansatz. In fact, since we are interested in a strictly bosonic theory where the  $B$-field is a geometric structure, for instance as a K\"{a}hler two form, hence the form of dilaton field is introduced merely
by solving the $\beta$-functions.  From Einstein-frame point of view, the solution of one-loop $\beta$-functions is  an effective gravitational theory solution. However, by starting  from a $\sigma$-model,   not directly from the effective action, the origin of arising   complex structure in the effective action becomes more clear. Furthermore, in the $\sigma$-model approach there is the advantageous of the $T$-duality transformation  for constructing new classes of equivalent solution. As mentioned before, in fact the solution of lowest order of $\beta$-functions dose not provide an exactly conformal $\sigma$-model.

\section{A static solution of the effective gravitational equations at one-loop in $(1+4)$-dimension}
This section is devoted to find a solution of the $\beta$-function equations \eqref{beta1}-\eqref{beta3} in a special case in which the $B$-field includes a complex structure. Before attempting to do this, let us remark some requirements. We are interested in an integrable complex structure, i.e. vanishing Nijenhus tensor, with $\nabla J\neq0$.

In the particular dimension of $(1+4)$, we take a general metric ansatz (in string frame) of the form
\begin{eqnarray}\label{gm}
d\hat{s}^{2}=g_{\hat{\mu}\hat{\nu}}dx^{\hat{\mu}}dx^{\hat{\nu}}&\equiv&- N\left( t \right) ~dt^2+g_{ij}dx^{i}dx^{j}\nonumber\\
&=&- N\left( t \right) ~dt^2+\frac{a^{2}\left( t \right)}{1-kr^2}dr^2+b^{2}\left( t \right)r^2 (d\theta^2+\sin^2{\theta}d\varphi^2)+ c^{2}(t)  ~dx^{2},
\end{eqnarray}
where
$$
\{\hat{\mu}\}=\{0,1,2,3,4\},~ \{i\}=\{1,2,3,4\},
$$
and $x^{4}\equiv x$ being an extra space-like dimension.

The $4$-dimensional space will be required to be a complex manifold carrying a complex structure $J$, and in this context the $g_{ij}$ is to be the associated Hermitian metric. So, the first step toward finding a solution for the $\beta$-function equations \eqref{beta1}-\eqref{beta3} with the proposed $B$-field in \eqref{b} is to determine the $\Omega_{ij}$ (or equivalently the $J_{i}^{~j}$). Thus, we must solve the equations of complex structure \eqref{cs} and Hermitian metric condition \eqref{hm} along with the integrability condition, $N_{ij}^{k}=0$, in \eqref{N}. Solving them with the above metric ansatz gives the following Hermitian structure and $B$-field
\begin{eqnarray}
J&=&J_{i}^{~j}dx^{i}\otimes\partial_{j}\nonumber\\
&=&\frac{c(t)\sqrt { \left( 1-k{r}^{2} \right)}}{a(t)}dx\otimes\partial_{r}+\frac{a(t)}{c(t)\sqrt { \left( 1-k{r}^{2} \right)}}dr\otimes\partial_{x}-\frac{1}{\sin{\theta}}d\theta\otimes \partial_{\varphi}-\sin{\theta}d\varphi\otimes \partial_{\theta},
\end{eqnarray}
\begin{eqnarray}\label{gb}
\hat{B}=\frac{\gamma}{2}(-\sqrt { \left( -k{r}^{2}+1 \right) ^{-1}}a \left( t \right) c
\left( t \right)dr\wedge dx+ b\left( t \right)^{2}{r}^{2}\sin{\theta}     d\theta\wedge d\varphi
).
\end{eqnarray}
The predicted time dependent functions $a(t)$, $b(t)$ and $c(t)$ in the above structure will be fixed by solving the $\beta$-function equations. According to \eqref{gb}, the non-zero components of field strength tensor $H_{\hat{\mu}\hat{\nu}\hat{\rho}}$ can be found as follows
\begin{eqnarray}
\hat{H}_{trx}=-\gamma\,\sqrt { \left(1- k{r}^{2} \right) ^{-1}} \left( \dot{a}(t) b \left( t
\right) +a\left( t \right)  \dot{b} \left( t
\right)  \right) ,
\end{eqnarray}
\begin{eqnarray}
\hat{H}_{r\theta\varphi}=-2\,\gamma\, c^2\left( t
\right)r\,\sin{\theta}  ,
\end{eqnarray}
\begin{eqnarray}
\hat{H}_{t\theta\varphi}=-2\,\gamma\, c\left( t
\right) \dot{c} \left( t
\right){r}^{2}\sin \theta.
\end{eqnarray}
The Weyl anomaly coefficients, the $\beta$-function equations of the above metric and $B$-field with a consistent dilaton field $\phi$ which is a function of $(t,r)$ are given in the Appendix.

After solving the $\beta$-function equations, we get two set of solution where all the scale factor functions which are anticipated in space part of the metric \eqref{gm} are fixed as constants with the following relations
\begin{eqnarray}
a(t)={\frac {\sqrt {3}}{{3\it q_{1}}}},~~b(t)=\sqrt {3}{\it q_{1}},~~c(t)=q_{1},
\end{eqnarray}
where the $q_{1}$ is an arbitrary constant and the other function and parameters are given as following two type

i:
\begin{eqnarray}\label{a}
k=0,~~\gamma=\frac{\sqrt{2}}{2},~~N(t)={\frac {\dot{F}(t)^2}{\Lambda}}, ~~\hat{\phi}=\ln  (r) +{\it F} \left( t
\right)+\phi_{0},
\end{eqnarray}
where $F(t)$ is an arbitrary function of time, $\Lambda$ is a positive parameter and $\phi_{0}$ is constant part of dilaton and

ii:
\begin{eqnarray}\label{a}
k=0,~~\gamma=\frac{\sqrt{2}}{2},~~\hat{\phi}=\ln  (r)+\phi_{0} ,~~~\Lambda=0.
\end{eqnarray}
where the lapse function $N(t)$ remains arbitrary in the second solution .

The dilatons in these solutions are  logarithmic function of $r$.  This type of dilaton field specially appears in some black hole solution in string theory for example \cite{dilaton(r)}. As mentioned before, if the $\Lambda$ is a result of charge deficit in the theory, the higher order corrections of $\beta$-functions must be considered. But, the first set of solutions which have a non zero  $\Lambda$ is not consistent with two-loop  $\beta$-functions. On the other hand,  the  $\Lambda$ in critical dimensionality can be a result of flux or curvature in the remaining dimensions. But, this regime requires the cosmological constant (minimum of the potential) to be zero or negative. However, the first solution is not allowed again and the only accepted solution will be the second one with a zero  $\Lambda$.

Consequently, the explicit form of $(1+4)$-dimensional backgrounds achieved by considering a complex structure $J$ in  the space part of $B$-field, has the following form
\begin{eqnarray}\label{5m}
d\hat{s}^{2}_{string frame}=-N(t)~dt^2+{{\it q_{1}}}^{2}(3dr^2+r^2 d\theta^2+r^2 \sin^2{\theta}d\varphi^2)+{\frac {{\it 1}}{{3{\it q_{1}}}^
                {2}}}~dx^{2},
\end{eqnarray}
\begin{eqnarray}\label{5b}
\hat{B}=\frac{\sqrt {2}}{4}({{\it q_{1}}}^{2}{r}^{2}\sin  \theta
d\theta\wedge d\varphi+dr\wedge dx).
\end{eqnarray}

\section{Dimension reduction to $(1+3)$-dimensions}
In this section, we are going to reduce the dimension of the $(1+4)$  theory (with coordinates $(t,r,\theta,\varphi,x)$) to the  $(1+3)$-dimensional space-time (with coordinates $(t,r,\theta,\varphi)$). By using the standard technique of Scherk and Schwarz \cite{Schwarz} one can parameterize the $5$-bein as follows \cite{4d}
\begin{eqnarray}
{\hat{e}}_{{\hat{\mu}}}^{~\hat{a}}=\left[ \begin {array}{cc} {\it e_{\mu}^{~a}}&{\it qA_{\mu}}\\ \noalign{\medskip}0&q
\end {array} \right].
\end{eqnarray}
Here, it is assumed that the fields are independent of the $x$ coordinate, i.e. there is a killing vector ${\hat{q}}_{\hat{\mu}}$ defined by \cite{4d}
\begin{eqnarray}
{\hat{q}}^{\hat{\mu}}\partial_{\hat{\mu}}=\partial_{\underline{x}},
\end{eqnarray}
where $\underline{x}$ is the flat version of the $x$ coordinate and $q=\sqrt{{\hat{q}}_{\hat{\mu}}{\hat{q}}^{\hat{\mu}}}$. If the $\mu$ and $\nu$ indices run over $\{0,1,2,3\}$ then $(1+4)$-dimensional fields will be decomposed as following
\begin{eqnarray}
\hat{g}_{xx}&=&\eta_{xx}q^2,~~\hat{B}_{x\mu}=B_{\mu},\nonumber\\
\hat{g}_{x\mu}&=&\eta_{xx}q^2A_{\mu},~~\hat{B}_{\mu\nu}=B_{\mu\nu}+A_{[\mu}B_{\nu]},\nonumber\\
\hat{g}_{\mu\nu}&=&g_{\mu\nu}+\eta_{xx}q^2A_{\mu}A_{\nu},~~\hat{\phi}=\phi+\frac{1}{2}\ln{q},
\end{eqnarray}
where the $\{g_{\mu\nu},B_{\mu\nu},\phi,A_{\mu},B_{\mu},q\}$ are $(1+3)$-dimensional fields. So, the expressions of them  in terms of $(1+4)$-dimensional fields are
written as follows
\begin{eqnarray}\label{dr}
g_{\mu\nu}&=&\hat{g}_{\mu\nu}-\frac{\hat{g}_{x\mu}\hat{g}_{x\nu}}{\hat{g}_{xx}},~~\phi=\hat{\phi}-\frac{1}{4}\ln{|\hat{g}_{xx}|},\nonumber\\
B_{\mu\nu}&=&\hat{B}_{\mu\nu}+\frac{\hat{g}_{x[\mu}\hat{B}_{\nu] x}}{\hat{g}_{xx}},~~B_{\mu}=\hat{B}_{x\mu},\nonumber\\
A_{\mu}&=&\frac{\hat{g}_{x\mu}}{\hat{g}_{xx}},~~q=|\hat{g}_{xx}|^{1/2}.
\end{eqnarray}
Applying these on the  metric, $B$-field and dilaton in \eqref{5m}, \eqref{5b} and \eqref{a} we get the background fields as
\begin{eqnarray}\label{4m}
ds^{2}_{string frame}=-N(t)~dt^2+{{\it q_{1}}}^{2}(3dr^2+r^2 d\theta^2+r^2 \sin^2{\theta}d\varphi^2),
\end{eqnarray}
\begin{eqnarray}\label{4fi}
\phi=\ln  \left( r \right) +\frac{1}{4}\ln
\left(3q_{1}^2\right)+\phi_{0},
\end{eqnarray}
\begin{eqnarray}\label{4b}
B_{\theta\varphi}=\frac{\sqrt {2}}{2}{{\it q_{1}}}^{2}{r}^{2}\sin\theta
,~~B_{\mu}=\frac{\sqrt{2}}{2},
\end{eqnarray}
\begin{eqnarray}
A_{\mu}=0,~~~~q=\frac{\sqrt{3}}{3q_{1}}.
\end{eqnarray}
In turn, the $(1+3)$-dimensional string effective action will contain the same terms as in \eqref{effective} in which the $(1+4)$-dimensional fields are replaced by $(1+3)$-dimensional ones \eqref{4m}-\eqref{4fi}.

In general,   by a time rescaling $dt'=\sqrt{N(t)} dt$  in   the metric \eqref{4m}, the arbitrary $N(t)$ function could be absorbed and  the metric recast in the following form
\begin{eqnarray}
d\hat{s}^{2}_{string frame}=-dt'^2+{{\it q_{1}}}^{2}(3dr^2+r^2 d\theta^2+r^2 \sin^2{\theta}d\varphi^2).
\end{eqnarray}
We will use this rescaled metric in the following.

In order to study the above result in the Einstein frame we may use a conformal transformation to obtain the metric in Einstein frame. So, in $(1+3)$-dimensions
\begin{eqnarray}
ds^{2}_{Einstein frame}=e^{\phi}ds^{2}_{string frame},
\end{eqnarray}
or equivalently
\begin{eqnarray}
\tilde{g}_{\mu\nu}=e^{\phi}g_{\mu\nu},
\end{eqnarray}
whose Ricci scalar is
\begin{eqnarray}
\tilde{R}={\frac {{3}^{3/4}e^{-\phi_{0}}}{18{{\it q_{1}}}^{5/2}{r}^{3}}}.
\end{eqnarray}
Obviously, in the limit $r\longrightarrow \infty$ we have $\tilde{R}\longrightarrow 0$, which accounts for the asymptotically flatness. The corresponding
Kretschmann scalar is obtained as
\begin{eqnarray}
\tilde{K}={\frac {31\,\sqrt {3}e^{2\phi_{0}}}{36\,{{\it  q_{1}}}^{3}{r}^{2}}},
\end{eqnarray}
which shows an essential singularity at $r\longrightarrow 0$. This singularity will be reconsidered in the following.

Now, we consider the Einstein equations \eqref{Ee} for this space-time in order to construct an effective matter source. The Einstein equations are as follows
\begin{eqnarray}
\tilde{R}_{\mu\nu}-\frac{1}{2}\tilde{R}\tilde{g}_{\mu\nu}=\kappa^2(T^{(\phi)}_{\mu\nu}+T^{(H)}_{\mu\nu}),
\end{eqnarray}
where $T^{(\phi)}_{\mu\nu}$ and $T^{(H)}_{\mu\nu}$ are given by (10) and (11).
So, the corresponding  energy-momentum tensor (as a perfect fluid) is given by\footnote{The $e_{a}^{~\mu}$ in non-coordinate basis are satisfying the relation
        $$
        g_{\mu\nu}e_{a}^{~\mu}e_{a}^{~\nu}=\eta_{ab},~~ \eta_{ab}=diag(-1,1,1,1).
        $$
        In the metric \eqref{4m} we have $e_{a}^{~\mu}$  as $e_{0}^{~0}=1,e_{1}^{~1}=\frac{\sqrt{3}}{3q_{1}}, e_{2}^{~2}=\frac{1}{q_{1}r}, e_{3}^{~3}=\frac{1}{q_{1}r sin\theta}$.}
\begin{eqnarray}\label{31}
T_{\mu\nu}=\kappa^2(T^{(\phi)}_{\mu\nu}+T^{(H)}_{\mu\nu})=\rho(r) ~e^{0}_{~\mu} e^{0}_{~\nu}+ P_{r}(r) ~e^{1}_{~\mu} e^{1}_{~\nu}+P_{l}(r)~ (e^{2}_{~\mu} e^{2}_{~\nu}+ ~e^{3}_{~\mu} e^{3}_{~\nu}),
\end{eqnarray}
where $P_{r}(r)$ and $P_{l}(r)$ are radial and lateral pressures. In this way one can obtain
\begin{eqnarray}
\rho(r)=P_{1}(r)=3P_{2}(r)={\frac {1}{4{q_{1}}^{2}{r}^{2}}}.
\end{eqnarray}
This energy-momentum tensor represents an anisotropic inhomogeneous barotropic  matter with  a positive $q_{1}$. We know that the matter obeys the barotropic equations of state as follows
\begin{equation}
P_{r}(r)=w_{r}\rho(r), ~~P_{l}(r)=w_{l}\rho(r),
\end{equation}
such that the two positive constants are given by $w_{r}=1$ and $w_{l}=\frac{1}{3}$.

Note that there are divergences in the pressures and Ricci scalar at $r=0$ which may be interpreted by the presence of an effective charge introduced by the $B$-field and dilaton at $r=0$\footnote{In fact, this divergence is appeared in both $B$-field and dilaton energy momentum tensor. The divergence which is caused by the dilaton is originally the result of the particular form of dilaton filed as a logarithmic function of $r$ \eqref{4fi} in this model. }. In other words, in this example the geometry is so deformed that there is effectively a charge at $r=0$. The $H_{t\nu\rho}$ is zero and  the only non-zero component of field strength $H_{\mu\nu\rho}$ is the  $H_{r\theta\varphi}$, hence the nature of this charge is magnetic and not electric\footnote{The Kalb-Ramond charge density vector which is visualized by electric charge is given by $Q^{\nu}=\partial_{\mu}H^{t\mu\nu}$\cite{zw}. This charge is zero becouse of $H_{t\nu\rho}=0$. }. The total magnetic charge associated with $H$ is calculated in Einstein-frame as follows
\begin{eqnarray}
M=\int e^{\phi}H=3^{1/4}q_{1}^{5/2}e^{\phi_{0}}\int r^2 \sin(\theta)~dr\wedge d\theta \wedge d\varphi.
\end{eqnarray}
Therefore, the magnetic charge per volume unit is  $g=3^{1/4}q_{1}^{5/2}e^{\phi_{0}}$and
hence the $q_{1}$ parameter in the metric is related to a magnetic charge.

It is worth mentioning the energy conditions. One can show that the perfect fluid (43) simply satisfies all the energy conditions, including
the null energy condition (NEC) $\rho+P_{i}\geq0$,
the strong energy condition (SEC) $
\rho+\Sigma_{i}P_{i}\geq0, ~\rho+P_{i}\geq0$,  the dominant energy condition (DEC) $
\rho\geq0,~\mid{\rho}\mid\geq \mid P_{i}\mid
$ and the weak energy condition (WEC) $\rho\geq0, ~~\rho+P_{i}>0$.

\section{$T$-dual solutions}
$T$-duality (target space duality) \cite{tduality} can be generally used in the $\sigma$-model context to generate a new class of solutions and background. The presented  example at the previous section is independent of the two $x$ and $\varphi$ coordinates, which will be regarded as isometry coordinates. In this section, we investigate the $T$-duality in $(1+4)$ and $(1+3)$-dimensions.

In $(1+4)$-dimensions, Buscher's $T$-duality transformations \cite{tduality} with respect to the isometry direction of $x$ have the following form\footnote{We are using the notion of $\beta$-function equations of \eqref{beta1}-\eqref{beta3} where the dilaton is minus twice  of the dilaton of the Ref. \cite{beta} and \cite{tduality}. Therefor, the $T$-duality transformation of dilaton of \cite{tduality} with the form of $\bar{\hat{\phi}}=\hat{\phi}-1/2 \ln|\hat{g}_{xx}|$ is matched with this coefficient.}:
\begin{eqnarray}\label{tdual5}
\bar{\hat{g}}_{xx}=\frac{1}{\hat{g}_{xx}},\bar{\hat{g}}_{x\mu}=\frac{\hat{B}_{{x\mu}}}{\hat{g}_{xx}},\bar{\hat{B}}_{x\mu}=\frac{\hat{g}_{{x\mu}}}{\hat{g}_{xx}},\nonumber\\
\bar{\hat{B}}_{\mu\nu}=\hat{B}_{\mu\nu}+\frac{(\hat{g}_{x\mu}\hat{B}_{{\nu x}}-\hat{g}_{x\nu}\hat{B}_{{\mu x}})}{\hat{g}_{xx}},\nonumber\\
\bar{\hat{g}}_{\mu\nu}=\hat{g}_{\mu\nu}-\frac{(\hat{g}_{x\mu}\hat{g}_{{x\nu}}
        -\hat{B}_{{x}\mu}\hat{B}_{x\nu})}{\hat{g}_{xx}},\nonumber\\
\bar{\hat{\phi}}=\hat{\phi}+ \ln{|\hat{g}_{xx}|}.
\end{eqnarray}
where $\bar{\hat{g}}_{\hat{\mu}\bar{\nu}}$, $\bar{\hat{B}}_{\hat{\mu}\hat{\nu}}$ and $\bar{\hat{\phi}}$ are metric, antisymmetric tensor and dilaton fields of the following dual $\sigma$-model
\begin{eqnarray}
I=\frac{1}{4\pi\alpha '}\int_{\Sigma}{\sqrt{h}d^2z(h^{\alpha\beta}\bar{\hat{g}}_{\hat{\mu}\hat{\nu}}\partial_{\alpha}{X}^{\hat{\mu}}\partial_{\beta}{X}^{\hat{\nu}}+\epsilon^{\alpha\beta}\bar{\hat{B}}_{\hat{\mu}\hat{\nu}}\partial_{\alpha}{X}^{\hat{\mu}}\partial_{\beta}{X}^{\hat{\nu}}+\alpha ' R^{(2)}\bar{\hat{\phi}})}.
\end{eqnarray}
For the rescaled of model \eqref{5m}, \eqref{5b} after using transformation \eqref{tdual5} one can find the following dual metric
\begin{eqnarray}
d\bar{\hat{s}}^{2}_{string frame}&=&\bar{\hat{g}}_{\hat{\mu}{\hat{\nu}}}dx^{\hat{\mu}}dx^{\hat{\nu}}\nonumber\\
&=&-dt^2+{{\it q_{1}}}^{2}(\frac{9}{2}dr^2+r^2 d\theta^2+r^2 \sin^2{\theta}d\varphi^2+3~dx^{2}+\frac{3\sqrt{2}}{2}drdx),
\end{eqnarray}
and dual antisymmetric and dilaton fields
\begin{eqnarray}
\bar{\hat{B}}=\frac{\sqrt {2}}{4}{{\it q_{1}}}^{2}{r}^{2}\sin \left( \theta
\right) d\theta\wedge d\varphi,
\end{eqnarray}
\begin{eqnarray}
\bar{\hat{\phi}}=\ln  \left( r \right) +\frac{1}{2}\ln  \left( 3{\it q_{1}}^2 \right) +\phi_{0}.
\end{eqnarray}

The other isometry coordinate is the $\varphi$ coordinate. Similar to the \eqref{tdual5}, the $T$-dual transformation with respect to $\varphi$ in $(1+4)$-dimensions gives the following $T$-dual metric, antisymmetric and dilaton fields respectively as follows
\begin{eqnarray}\label{8}
d\bar{\hat{s}}^{2}_{string frame}&=&\bar{\hat{g}}_{\hat{\mu}{\hat{\nu}}}dx^{\hat{\mu}}dx^{\hat{\nu}}\nonumber\\
&=&-dt^2+3{{\it q_{1}}}^{2}dr^2+\frac{3}{2}{{\it q_{1}}}^{2}r^2 d\theta^2+\frac{1}{{{\it q_{1}}}^{2}r^2 \sin^2{\theta}}d\varphi^2+\frac{\sqrt{2}}{2sin \theta}d\theta d\varphi+\frac{1}{3{{\it q_{1}}}^{2}}~dx^{2},
\end{eqnarray}
\begin{eqnarray}\label{9}
\bar{\hat{B}}=\frac{\sqrt {2}}{4}dx\wedge dr,
\end{eqnarray}
\begin{eqnarray}\label{10}
\bar{\hat{\phi}}=3\ln  \left( r \right) +2\,\ln
\left( \sin{\theta}  \right)+\phi_{0} +2\ln  \left( {\it q_{1}} \right).
\end{eqnarray}

In $(1+3)$-dimension (32)-(35), the only isometry coordinate is $\varphi$.
We may apply two procedures here: we can find the $T$-dual backgrounds in $(1+4)$-dimensions and then dimensionally reduce the $x$ coordinate; or, we may reduce the $x$ coordinate and then find the $T$-dual solution with respect to $\varphi$.
In the second procedure, after dimension reduction of $x$ coordinate,
Buscher's $T$-duality transformation with respect to $\varphi$ coordinate are given as follows
\begin{eqnarray}
\bar{g}_{\varphi\varphi}=\frac{1}{g_{\varphi\varphi}},\bar{g}_{\varphi m}=\frac{B_{{\varphi m}}}{g_{\varphi\varphi}},\bar{B}_{\varphi m}=\frac{g_{{\varphi\mu}}}{g_{\varphi\varphi}},\nonumber\\
\bar{B}_{mn}=B_{mn}+\frac{(g_{\varphi m}B_{{n\varphi}}-g_{\varphi n}B_{{m\varphi}})}{g_{\varphi\varphi}},\nonumber\\
\bar{g}_{mn}=g_{mn}-\frac{(g_{\varphi m}g_{{\varphi n}}
        -B_{{\varphi}m}B_{\varphi n})}{g_{\varphi\varphi}},\nonumber\\
\bar{\phi}=\phi+ \ln{|g_{\varphi\varphi}|}.
\end{eqnarray}
Here, $m$ and $n$ indices indicate the $\{t,r,\theta\}$ coordinates. Then using \eqref{4m}-\eqref{4fi}, the $T$-dual answers with respect to $\varphi$ coordinate are given as follows
\begin{eqnarray}\label{11}
d\bar{s}^2 _{\{string frame\}}&=&\bar{g}_{\mu\nu}dx^{\mu}dx^{\nu}\nonumber\\
&=&-dt^2+3q_{1}^2dr^2+\frac{3}{2}q_{1}^2 r^2 d\theta ^{2}+\frac{1}{q_{1}^2 r^2  \sin^2 \left( \theta \right)}d\varphi^2+\frac{\sqrt{2}}{ 2\sin\theta}d\theta d\varphi,
\end{eqnarray}
\begin{eqnarray}\label{13}
\bar{\phi}=3\ln  \left( r \right)  +2\,\ln
\left( \sin{\theta}  \right) +\phi_{0} +1/4\,\ln
\left( 3 \right) +5/2\,\ln  \left( {\it q_{1}} \right) ,
\end{eqnarray}
such that the $T$-dual $B$-field is zero. On the other hand, if we apply the dimension reduction according to \eqref{dr} on the $T$-dual solutions \eqref{8}-\eqref{10}, we will obtain the same results as \eqref{11}, \eqref{13} and similar to it, the reduced $B$-field would be zero, i.e. \textit{the dimension reduction procedure and $T$-duality transformation are consistent with each other}.
\section{Conclusion}
We have developed a $\sigma$-model on a manifold $R\times N$ whose space part $N$ is a complex manifold. Specially, following the idea of giving a physical implication  to the complex structure, we have considered the complex structure of the complex manifold  as the antisymmetric tensor field of the $\sigma$-model in the form of $\hat{B}= \frac{\gamma}{2}~ g_{kj}J_{i}^{~k}dx^{i}\wedge dx^{j}$. Therefor, since the metric field in $\sigma$-model is a geometrical structure, the role of antisymmetric filed is played by a geometrical structure on the manifold.
Two static solutions are found for the one-loop $\beta$-functions in the $(1+4)$-dimensions, one with a non-zero and positive  $\Lambda$ and another with a zero $\Lambda$. Considering the nature of $\Lambda$ as a charge deficit in non-critical dimensions, the second solution is preferred.
Then, its dimension reduction  to  $(1+3)$-dimensional  space-time is performed
and resulted in a static solution of Einstein equation with the energy-momentum tensor corresponding to an anisotropic inhomogeneous barotropic matter which satisfies all the energy conditions. From gravitational point of view, this
is interpreted as a static solution of an effective gravity theory with a dilaton and antisymmetric $B$-field representing matter sources.  There is a divergence in the energy Ricci scalar, density and pressures at $r=0$ which is interpreted as a natural result of  the presence of an effective magnetic charge  at $r=0$. Finally, $T$-dual solutions are constructed in both $(1+4)$ and  $(1+3)$-dimensions. We have shown that in  $(1+3)$-dimensional space-time the $T$-duality transformation and the dimension reduction procedures are compatible.

\section*{Acknowledgment}
We would like to express our sincere  gratitude to M. M. Sheikh-Jabbari  for his useful comments. This research has been supported by Azarbaijan Shahid Madani university by a research fund No. 401.231.
\section*{Appendix}
Here we give the non zero components of the $\beta$-functions $\beta_{\hat{\mu}\hat{\nu}}$ related to the metric given by \eqref{gm} and the B-field given by \eqref{gb}, as follows \footnote{The overdot and prime stand for diffrentiation $\frac{\partial}{\partial t}$ and $\frac{\partial}{\partial r}$, respectively.}
\begin{eqnarray}
&&\hspace{-20mm}\beta_{00}^{g}=- \{ [\dot{\phi} \left( t,r \right)+\dot{\ln}{a}(t)+2\,\dot{\ln}{b}(t) +\dot{\ln}{c}(t)  ] \dot{N} \left( t
\right) -{\gamma}^{2}N (t)  [(\dot{\ln}{a}(t)+\dot{\ln}{c}(t) ) ^{2} +4\,  \dot{\ln}{b}(t)  ^{2}]
\} a (t)   ^{2}  c \left( t
\right)   ^{2}  b (t)  ^{2}
\nonumber\\
&&\hspace{-6mm}+ 2\, c (t)   ^{2}N (t) a
(t) b (t)  [ \ddot{\phi} \left( t,r \right) a \left( t
\right) b (t) +2\,a (t) \ddot{b }(t) +b (t)\ddot{a} (t)  ]+2\,c (t)  a (t)   ^{2} \ddot{c} (t)
b (t) ^{2}N (t) ,
\end{eqnarray}
\begin{eqnarray}
&&\hspace{-13mm}\beta_{11}^{g}=2\,c (t) ^{2}b \left( t
\right)N (t) \{{r}^{2}[ N (t)  \left( k{r}^{2}-1
\right) \phi'' \left( t,r
\right) +kr \phi' \left( t,r \right) N (t) +1/2\, \dot{a} (t) ^{2}{\gamma}^{2}+a \left( t
\right)  \dot{a} (t)
\dot{\phi} \left( t,r \right) +a \left( t
\right) \ddot{a} (t)
]
\nonumber\\
&&\hspace{-5mm}~ +2\, [ a (t) {r}^{2}  \dot{a} (t)   ( \frac{1}{2} \left( {
        \gamma}^{2}+1 \right)\dot{\ln}{c}( t) +\dot{\ln}{b}(t) -\frac{1}{4}\dot{\ln}{N}( t)  ) +\frac{1}{4}
\dot{\ln}{c}( t) ^{
        2}{r}^{2}{\gamma}^{2}  a (t)  ^{2}+N
(t)  \left( k \left( {\gamma}^{2}+1 \right) {r}^{2}-{
        \gamma}^{2} \right) ]\},\nonumber\\
\end{eqnarray}
\begin{eqnarray}
&&\hspace{-58mm}\beta_{10}^{g}=  r\phi'' \left( t,r
\right)  b \left( t \right) a \left( t \right) -\dot{a}\left( t \right)  \phi' \left( t,r \right)   b \left( t
\right) r-2\, \dot{a}\left( t \right)b \left( t \right) +2\, \dot{b}\left( t \right) a \left( t \right)  \left( {\gamma}^{2}+1
\right) ,
\end{eqnarray}
\begin{eqnarray}
&&\hspace{-11mm}\beta_{22}^{g}=N (t) c (t) [ b (t)
^{2} r\{ \left( k{r}^{2}-1 \right) N (t)  \phi' \left( t,r \right) + \dot{\ln}{b}( t) \left(
\,\dot{\phi} \left( t,r \right) + \dot{\ln}{b}( t)  \left( {
        2       \gamma}^{2}+1 \right)  \right) r  a (t)
^{2}\}+ a (t)  ^{2}b \left( t
\right)  \ddot{b} (t)  {r}^{2}]
\nonumber\\
&&\hspace{-3mm}+a \left(
t \right)  ^{2}c (t) N (t) b \left(
t \right) \dot{b} (t){r}^{2} [-\frac{1}{2}\dot{\ln}{N}( t)+\dot{\ln}{a}( t) +\dot{\ln}{c}( t) ]   +2\, [
( k ( {\gamma}^{2}+1 ) {r}^{2}-{\gamma}^{2}-\frac{1}{2}
)   b (t)^{2}+\frac{1}{2}  a
(t) ^{2} ]   N (t)
^{2}c (t),\nonumber\\
\end{eqnarray}
\begin{eqnarray}
&&\hspace{-153mm}\beta_{33}^{g}=1/2 \sin^2(\theta)\beta_{22}^{g},
\end{eqnarray}
\begin{eqnarray}
&&\hspace{-30mm}\beta_{44}^{g}=  a (t) ^{2}b (t) N( t) \{  {\gamma}^{2}  \dot{c}   (t)  ^{2}+2\,c (t)\dot{c} (t)   +2\, c (t)   \dot{
        c} (t)   [ \dot{\phi} \left( t,r \right)+  \left( {\gamma}^{2}+1 \right) \dot{\ln}{a} (t) +2\,\dot{\ln}{b} (t) -1/2\,\dot{\ln}{N} (t)  ]\} \nonumber\\
&&\hspace{-16mm}+c (t)  ^{2}b (t) N \left( t
\right)  \dot{a} (t)
^{2}{\gamma}^{2},
\end{eqnarray}
\begin{eqnarray}
&&\hspace{-34mm}\beta_{14}^{B}=2\,N (t) a (t) b (t)  \{
c (t)  ^{2}a (t)  \dot{\ln}{a} (t)^{2}-a
(t)  [ c (t)  ( \dot{\ln}{a} (t) +\dot{\ln}{c} (t) )  ]^{2}\dot{\phi} \left( t,r \right) -  c (t)  ^{2}\ddot{a} (t) -c \left( t
\right) a (t) \ddot{c}
(t)  \}
\nonumber\\
&&\hspace{-25mm}+c (t) N (t) a (t) b \left( t
\right)  \{a (t)  [\dot{\ln}{N} (t) -4\,\dot{\ln}{b} (t) +2\,\dot{\ln}{c} (t)
] \dot{c} (t) +c \left( t
\right)  \dot{a} (t)
( \dot{\ln}{N} (t) -4\,\dot{\ln}{b} (t)  )  \},
\nonumber\\
\end{eqnarray}
\begin{eqnarray}
&&\hspace{-94mm}\beta_{04}^{B}=\left(   \left( k{r}^{3}-r \right) \phi' \left(
t,r \right) -4\,k{r}^{2}+2
\right)  \left(
\dot{a} (t)c \left(
t \right) +a (t) \dot{c} (t))  \right),
\end{eqnarray}
\begin{eqnarray}
&&\hspace{-44mm}\beta_{23}^{B}= c (t)   b (t)  ^{2}N \left( t
\right)  \{{r}^{2} a (t) ^{2}
[-\frac{1}{2}\dot{\ln}{N} (t) +\dot{\ln}{a} (t) +\dot{\ln}{c} (t) ] \dot{\ln}{b} (t) +rN (t)  \left( k{r}^{2}-1 \right) \phi' \left( t,r \right) \nonumber\\
&&\hspace{-36mm}-  \dot{\ln}{b} (t)  ^{2}{r}^{2}
a (t) ^{2}+ \dot{\ln}{b} (t) {r}^{2}  a \left( t
\right)  ^{2}\dot{\phi} \left( t,r
\right) +N (t) \}  +\ddot{b} (t)
N (t) c (t) b (t)
a (t)  ^{2}{r}^{2},
\end{eqnarray}

\begin{eqnarray}
&&\hspace{-10mm}\beta^{\phi}=-4\,c (t) N (t) b (t) a \left(
t \right) {r}^{2} [ b (t) a (t) \ddot{c} (t) +c (t)
(   \ddot{a} \left( t
\right) b (t) +2\,a (t) \ddot{b} (t)  ) ]
\nonumber\\
&&\hspace{-2mm}-4\, [\left( k{r}^{2}-1 \right) N (t)  \phi'' \left( t,r \right) +  a
(t)  ^{2}\ddot{\phi} \left( t,r \right) +1/2N (t)  \left( k{r}^{2}-1
\right) \phi' \left( t,r
\right)  ^{2}+1/2 \dot{\phi} \left( t,r \right) ^{2}  a (t)
^{2} ] {r}^{2}  b (t) ^{2}
c (t)   ^{2}N (t)       \nonumber\\
&&\hspace{-2mm}-4\, N (t)   ^{2} c \left( t
\right) ^{2}\{ r b (t)  ^{2
} \left(3 k{r}^{2}-2 \right) \phi'
\left( t,r \right) + [ -1/2\,{r}^{2}\Lambda\,  a \left( t
\right) ^{2}+k \left( {\gamma}^{2}+1/3\, \right) {r}^{2}-{\gamma}^{2}-1] b (t)   ^{2}+1
/3\, a (t) ^{2} \}\nonumber\\
&&\hspace{-2mm}-8\,b (t)   c (t)   ^{2}
a (t) ^{2}N (t) {r}^{2}
\{-1/4\,b (t)  [ -2\,\dot{\ln}{a} (t)+\dot{\ln}{N} (t) -4\,\dot{\ln}{b} (t) -2\,\dot{\ln}{c} (t) ]\dot{\phi} \left( t,r \right) \nonumber\\
&&\hspace{-2mm}+1/8\,b \left( t
\right) \dot{\ln}{a} (t)^{2}{\gamma}^{2}+ [ 1/4\,b (t)  \left( {
        \gamma}^{2}+2 \right)\dot{\ln}{c} (t) -1/4\,b (t) \dot{\ln}{N} (t) +\dot{b} (t)
]\dot{\ln}{a} (t) \nonumber\\
&&\hspace{-2mm} -1/2\,b
(t)  [ -1/4\,\dot{\ln}{c} (t)^{2}{\gamma}^{2}+ [ 1/2\,\dot{\ln}{N} (t)-2\,\dot{\ln}{b} (t) ] \dot{\ln}{c} (t) + \dot{\ln}{b} (t)  [ \left( -{\gamma}^{2}-1 \right) \dot{\ln}{b} (t) +\dot{\ln}{N} (t) ]  ]\}.
\end{eqnarray}



\end{document}